\documentclass{article}
\usepackage{spconf,amsmath,graphicx}

\usepackage{enumitem}
\setlist{nosep, leftmargin=14pt}

\usepackage{mwe} 
\usepackage{amsthm}
\usepackage{algorithmicx,algorithm}
\usepackage{amsbsy}
\usepackage{amsfonts}
\usepackage{marvosym}
\usepackage{amsmath}
\usepackage{threeparttable}
\usepackage{booktabs}
\usepackage{multirow}
\usepackage{float}
\usepackage{xurl}
\usepackage{breakurl}

\title{active index: an integrated index to reveal disrupted brain network organizations of major depressive disorder patients}
\name{Anonymous Authors}
\address{Author Affiliation(s)}
\name{\begin{tabular}{c}Yu Fu$^{1,2}$\sthanks{These authors contributed equally to this work.}, Yanyan Huang$^{1\ast}$, Meng Niu$^{3\ast}$, Le Xue$^{1}$, Shunjie Dong$^{1}$,\\ Shunlin Guo$^{3}$, Junqiang Lei$^{3}$\sthanks{Corresponding authors.} and Cheng Zhuo$^{1\dagger}$\end{tabular}}

\address{$^{1}$ Zhejiang University, Hangzhou, China\\
$^{2}$ Binjiang Institute of Zhejiang University, Hangzhou, China\\
$^{3}$ Department of Radiology, The First Hospital of Lanzhou University, Lanzhou, China}

\begin{document}
%
\maketitle
\begin{abstract}

Altered functional brain networks have been a typical manifestation that distinguishes major depressive disorder (MDD) patients from healthy control (HC) subjects in functional magnetic resonance imaging (fMRI) studies. Recently, rich club and diverse club metrics have been proposed for network or network neuroscience analyses. The rich club defines a set of nodes that tend to be the hubs of specific communities, and the diverse club defines the nodes that span more communities and have edges diversely distributed across different communities. Considering the heterogeneity of rich clubs and diverse clubs, combining them and on the basis to derive a novel indicator may reveal new evidence of brain functional integration and separation, which might provide new insights into MDD. This study for the first time discussed the differences between MDD and HC using both rich club and diverse club metrics and found the complementarity of them in analyzing brain networks. Besides, a novel index, termed "active index", has been proposed in this study. The active index defines a group of nodes that tend to be diversely distributed across communities while avoiding being a hub of a community. Experimental results demonstrate the superiority of active index in analyzing MDD brain mechanisms.


\end{abstract}

\vspace{-0.5cm}
\section{Introduction}
\label{sec:intro}

Major depressive disorder (MDD), accompanied by alterations of brain structures and functions~\cite{yao2020morphological,yao2019structural}, is now one of the most prevalent mental illness worldwide~\cite{sagar2020burden,zhong2020prevalence}. As a non-invasive imaging technique, functional magnetic resonance imaging (fMRI) has become a major tool to promote the understanding of the principles and mechanisms underlying complex brain function changes caused by MDD. Recently, some novel graph theory metrics have been proposed and applied to the analyses of fMRI to reveal brain functional abnormalities associated with MDD at the height of network neuroscience~\cite{bassett2017network,bassett2018nature}. By applying graph metrics on whole-brain, previous studies have demonstrated losses of small-world characteristics, declines of nodal efficiencies~\cite{yao2019structural,yu2020abnormal}, variational local centralities~\cite{yun2021graph} and disrupted rich club properties~\cite{liu2021disrupted} in MDD group. From small-world characteristics to rich club metrics represents the increased understanding of information integration and functional modularity regarding the high-level structures of MDD brain.


Rich club refers to those nodes that have a disproportionately high number of edges as well as many edges between each other~\cite{bertolero2017diverse}, which usually be defined as inner-community hubs. Group-scale differences of rich club organizations (e.g., different ratios of hub nodes and non-hub nodes) usually suggest the disrupted brain functions in information transmission caused by disorders or diseases~\cite{liu2021disrupted,daianu2016disrupted,verhelst2018impaired}. However, as demonstrated by~\cite{bertolero2017diverse}, rich club nodes usually exist on the periphery of the brain network, which fails to describe these nodes that are more highly interconnected and their edges are more critical for efficient global integration. A recent study~\cite{bertolero2017diverse} proposed the diverse club to solve the above problem. Since rich club tends to be the hub of a specific community and diverse club appears to be in the topological center that bridges different network communities, we propose the active index, which takes the advantage of both rich club and diverse club, to represent those nodes tend to connect as many communities as possible, but avoiding being a hub of a specific community. Leveraging the three kinds of powerful hub metrics, this study aims to provide more critical evidence for the functional integration and segregation in both region and community scales of the MDD brain.

\vspace{-0.5cm}
\begin{table}[htbp]\footnotesize
 \centering
 \caption{Demographic and clinical characteristics of subjects.}
 \label{table1}
 \begin{threeparttable}
 \renewcommand\arraystretch{0.9} 
 \begin{tabular}{lllllll}  
  \toprule   
  \multicolumn{1}{c}{Variables} & MDD (n=71) & HC (n=71) & P-value  \\ 
  \midrule 
  Age (years) & 31.66$\pm$7.95 & 31.20$\pm$8.06 & 0.7298$^a$    \\
  Gender (M/F) & 28/43 & 29/42 & 0.809$^b$   \\
  HAMD & 24.69±4.77 & None & None   \\
  \bottomrule  
 \end{tabular}
 \begin{tablenotes}
    \item $^a$ Two-sample t-test. $^b$ Chi-square two-tailed test.
 \end{tablenotes}
 \end{threeparttable}
\end{table}

\section{Methology}
\label{sec:metho}

\begin{figure*}[htp]
 \centering
 \centerline{\includegraphics[width=14cm]{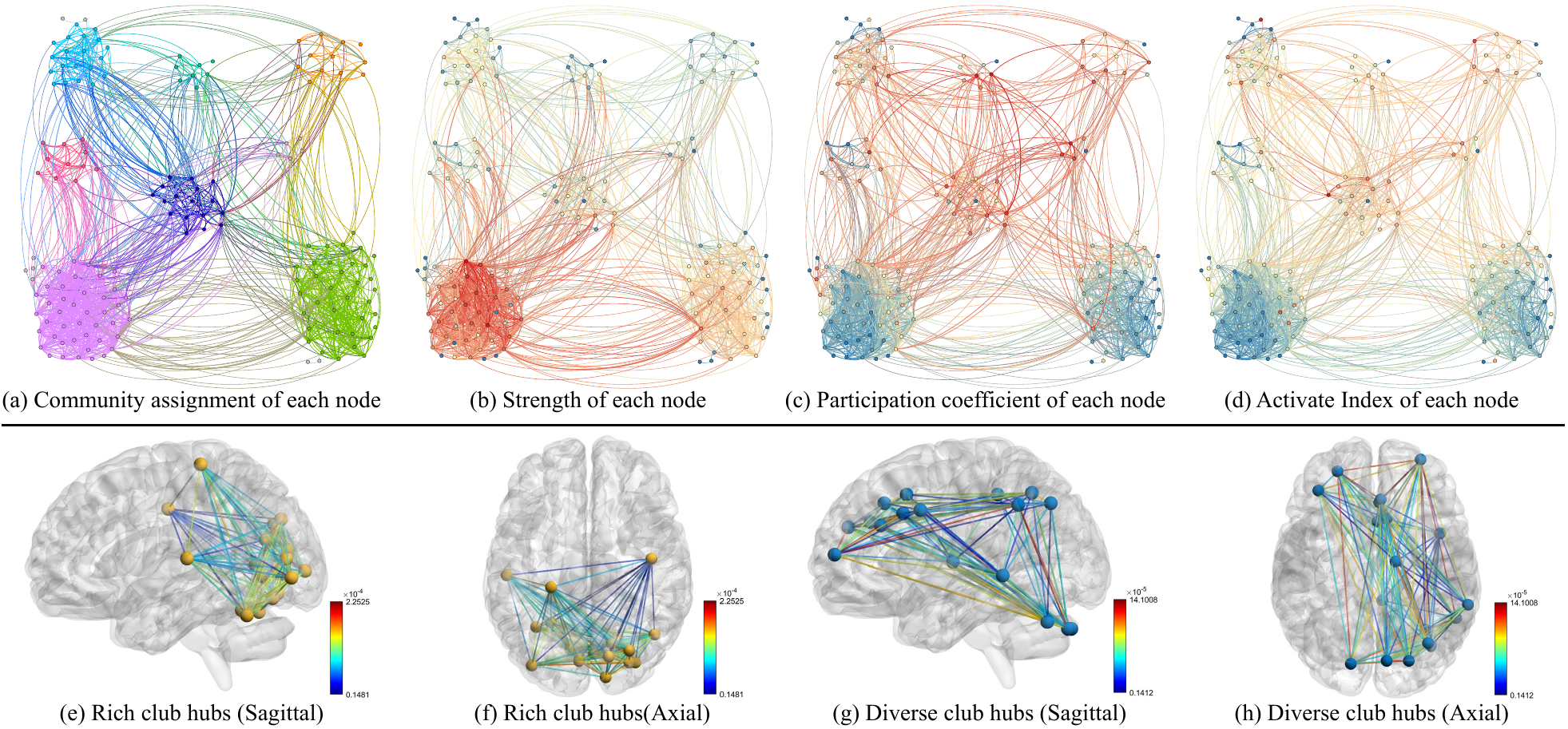}}
\caption{(a) The community assignment of each node in the human brain network using the InfoMap algorithm (different colors represent different communities). (b) The strength of each node; (c) The PC of each node; (d) The active index of each node. Nodes in red represent the maximum value for the given metric (strength, PC, or active index), yellow is medium, and blue is the minimum. Edges are colored by the mix between the two nodes each edge connects. (e) and (f) are the visualization results of rich club nodes in the brain. (g) and (h) are the visualization results of diverse club nodes in the brain.} 
\label{Network_clubnodes}
\end{figure*}

\subsection{Image acquisition and preprocessing}

All fMRI data adopted in this study were downloaded from a site of the REST-meta-MDD data~\footnote{http://rfmri.org/RfMRIMapsDataSharingStructure}, which contains 71 MDD patients and 71 age-matched HC subjects. Demographic and clinical characteristics of all subjects are summarized in Table~\ref{table1}. These fMRIs were preprocessed followed by a standardized preprocessing protocol on Data Processing Assistant for Resting-State fMRI (DPARSF), including: slice timing correction, realign, covariates removed, spatial normalization, and filter (0.01\textasciitilde0.1Hz).


\subsection{Construction of brain networks}

For each subject, the nodes (brain regions) of the network were demarcated to 160 functional regions according to the Dosenbach atlas~\cite{dosenbach2010prediction}. The edges were weighted by the strength of the Pearson correlations ($r$ values) between the two nodes’ time series of activity. The $r$ values between all pairs of signals were computed to form a $160\times160$ matrix, which was then Fisher z transformed. For each subject, we constructed weighted symmetric functional connectivity (FC) matrices over a wide range of network edge sparsities ($5\%$-$20\%$) with a step of 1$\%$ ~\cite{yao2019structural}. We have tried each choice of sparsity and finally the sparsity of 13$\%$ was mainly discussed in this study given the consideration of community detection and between-group analyses. The maximum spanning tree for each graph was calculated, and these edges were not removed in order to ensure that there are no isolated nodes in the graph ~\cite{bertolero2017diverse}.

The diverse club refers to a set of nodes with high participation coefficient (PC)~\cite{bertolero2017diverse}, and these nodes have edges diversely distributed across the whole network. Given a particular community assignment, the PC of of node $i$ can be defined as:

\begin{equation}
    \mathrm{PC}_{i}=1-\sum_{s=1}^{N_{\mathrm{M}}}\left(\frac{K_{i s}}{K_{i}}\right)^{2}
\end{equation}

\noindent where $K_{i}$ is the sum of $i$ 's edge weights, $K_{i s}$ is the sum of $i$ 's edge weights to community $s$, and $N_{\mathrm{M}}$ is the total number of communities. The community structure can be detected by InfoMap algorithm~\cite{rosvall2008maps}, which is based on how information randomly flows through the networks. Thus, the PC is a measure of how evenly distributed a node's edges are across communities. A node's PC is maximal if it has an equal sum of edge weights to each community in the network, and it becomes to 0 if all of its edges belong to a single community~\cite{bassett2017network}.

The rich club refers to a set of nodes with high strength and has a disproportionately high number of edges as well as many edges between each other~\cite{bertolero2017diverse,van2011rich}. These nodes are thought to be critical for global communication because of the high betweenness centrality of rich club members.


\begin{figure*}[htb]
 \centering
 \centerline{\includegraphics[width=12cm]{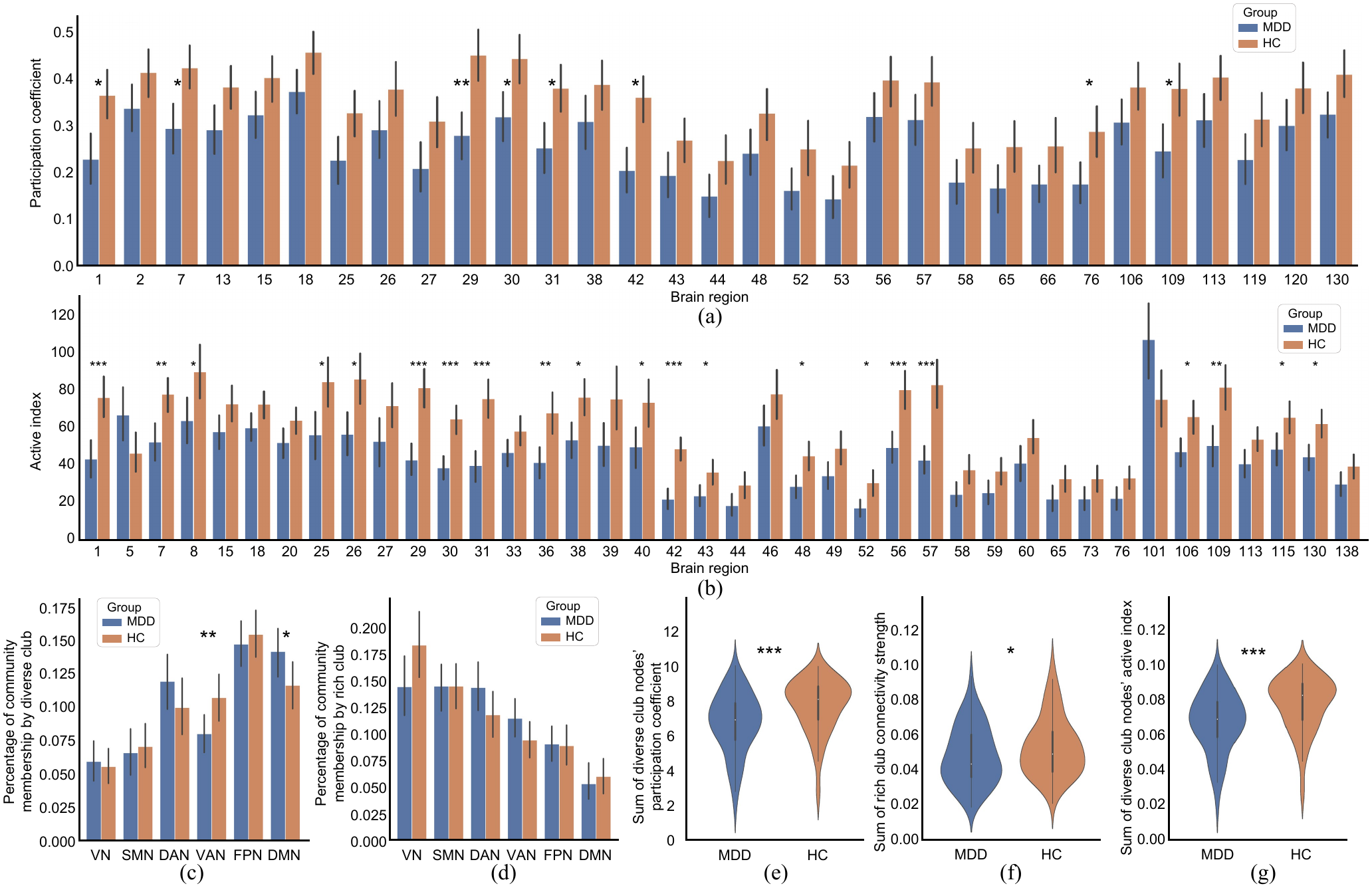}}
\caption{(a)-(b) Brain regions that have significant difference ($p < 0.05$) on PC and active index between MDD and HC groups, where the $^*$, $^{**}$ and $^{***}$ indicate this region survives the FDR correction under $p=0.05, t-test$, and represent $p<0.05$, $p<0.01$ and $p<0.001$ respectively. (c)-(d) The percentage of each human cognitive system that is comprised of nodes from each club in MDD and HC group, where $^*$ and $^{**}$ represent $p<0.05$ and $p<0.01$ in t-test, respectively. For ease of comparison, a canonical division of nodes into cognitive systems is used~\cite{yan2019reduced} (VN: visual network, SMN: sensory-motor network, DAN: dorsal attention network, VAN: ventral attention network, FPN: frontoparietal network, DMN: default mode network). (e) The distributions of the sum of the diverse club nodes' PCs in MDD and HC groups ($p=4.20\times 10^{-4}$). (f) The distributions of the sum of the rich club connectivity strength in MDD and HC groups ($p=0.038$). (g) The distributions of the sum of the diverse club nodes' active indexes in MDD and HC groups ($p=2.16\times 10^{-4}$).} 
\label{DiverseClubVsRichClub}
\end{figure*}

To be specific, clubs are defined on rank-ordering the nodes based on the PC or the strength and taking the nodes with a PC or strength above a particular rank. Club nodes are selected on the basis of the group-averaged network, which is constructed by thresholding the network at a particular cost, and in this study is $\text {cost} = 0.13$. Then, the brain network connections can be divided into 3 classes: club connections, feeder connections, and local connections~\cite{van2011rich}. Club connections link hub nodes; feeder connections link hub nodes and non-hub nodes; and local connections link non-hub nodes. Detailed comparisons of the three kinds of nodes can be found in our supplementary materials.




Together, our proposed active index can be given as follows:

\begin{equation}
    \alpha(i)=\frac{\mathrm{PC}_{i}}{\mathrm{Strength}_i}
\end{equation}

\noindent where $\alpha(i)$ denotes the active index of a specific brain region $i$, which can be easily calculated by the ratio of participation coefficient $\mathrm{PC}_{i}$ divided by node strength $\mathrm{Strength}_i$.

\section{Results}
\label{sec:results}

\subsection{Disrupted diverse club and rich club organizations in MDD group}

The detected brain community structure using the InfoMap algorithm is presented in Fig.~\ref{Network_clubnodes}(a). Both the strength and PC of each node in each brain network are presented in Fig.~\ref{Network_clubnodes}(b) and Fig.~\ref{Network_clubnodes}(c). We refer to the high ($90^{th}$ percentile and above) degree and PC nodes as rich club and diverse club respectively~\cite{bertolero2017diverse}. The whole-brain distributions of rich club nodes and diverse club nodes are visualized in Fig.~\ref{Network_clubnodes}(e)-(h). It can be concluded that rich club forms clusters on the periphery, which tends to be the hubs of specific communities; diverse club clusters in the center of the layout and have edges diversely distributed across different communities, which plays a more important role in brain functional integration.

We have compared the PCs between MDD and HC groups in 160 brain regions using t-test. Fig.~\ref{DiverseClubVsRichClub}(a) shows the brain regions that have significant difference ($p < 0.05$) in PC, where the X coordinate represents these regions' number defined in~\cite{dosenbach2010prediction}. There are totally 31 brain regions show significant difference between MDD and HC groups, and eight brain regions (denoted by *) survive the FDR correction at threshold of $p = 0.05$. Results of Fig.~\ref{DiverseClubVsRichClub}(a) suggest that PCs of MDD brain regions are significantly lower than that of HC brain regions in all between-group comparisons, which indicates the diverse club is disrupted in the MDD group. Fig.~\ref{DiverseClubVsRichClub}(c)-(d) show the percentage of community membership by both diverse club and rich club in six networks: visual network (VN), sensory-motor network (SMN), dorsal attention network (DAN), ventral attention network (VAN), frontoparietal network (FPN), and default mode network (DMN). In the percentage of community memberships by the diverse club, the VAN ($p=0.003$) and DMN ($p=0.032$) display significant differences between the MDD group and HC group, which partly supports the findings of~\cite{yan2019reduced} that directly using functional connectivity. Also, these results reveal that diverse club might work in a more sensitive manner than rich club in sub-network analyses of MDD patients.

Fig.~\ref{DiverseClubVsRichClub}(e)-(f) show the significant difference of MDD group vs. HC group in the sum of diverse club PC ($p = 4.20 \times 10^{-4}$) and in the sum of rich club connectivity strength ($p = 0.038$). These results indicate that the diverse club and rich club are both significantly disrupted in the MDD group, and the degree of disruption in the diverse club is more significant and sensitive. We also compare the differences in the club connections, feeder connections and local connections of both diverse club and rich club between MDD group and HC group, and details can be found in the supplementary materials.

\begin{figure}[htbp]
 \centering
 \centerline{\includegraphics[width=7cm]{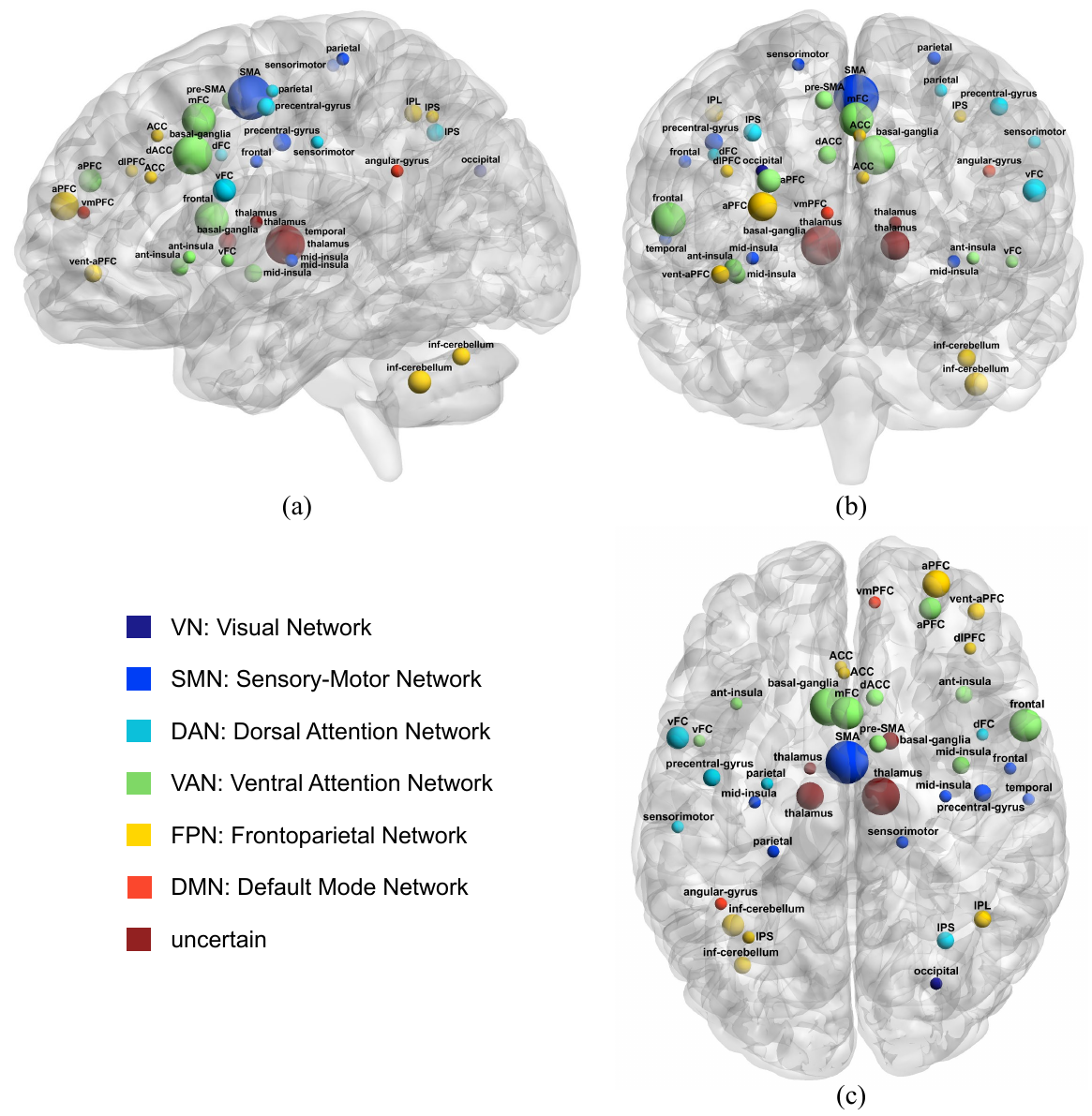}}
\caption{Brain regions that show significantly decreased active
indexes in the MDD group. Regions with the same color represent
they belong to the same functional sub-network according
to the Dosenbach atlas [13]. Regions with larger sizes
represent the altered active indexes in these regions are more
significant.} 
\label{activaIndexBrainNodes}
\end{figure}

\subsection{Distinguishing MDD patients from HC subjects by active index}

Based on Dosenbach atlas, 40 brain regions are demonstrated have significant differences ($p<0.05$) in the active index between MDD and HC groups (Fig.~\ref{DiverseClubVsRichClub}(b)), and there are totally 21 brain regions survive the threshold of $p = 0.05$ after FDR correction. It is easy to find that seven of the eight brain regions that survive the FDR correction in PC overlap with significant regions found by active index. However, active index can detect 13 more brain regions with significant between-group differences than PC. The sum of diverse club nodes' active index is illustrated in Fig.~\ref{DiverseClubVsRichClub}(g), which demonstrates the between-group difference based on active index is more significant ($p=2.16\times 10^{-4}$) than PC ($p = 4.20 \times 10^{-4}$). In conclusion, all evidence demonstrates that active index can distinguish MDD patients from HC subjects more effectively. We also display the between-group analysis results using other graph theory metrics, which can be found in our supplementary materials.

\subsection{Brain regions with significantly decreased active indexes in MDD group}

As illustrated in Fig.~\ref{activaIndexBrainNodes}, there are totally 40 brain regions (each ball represents one brain region) that have significantly decreased active index in MDD group. Most of them are located in 4 functional sub-networks include: SMN (8 regions), DAN (6 regions), VAN (10 regions) and FPN (9 regions) according to the Dosenbach atlas. The size of each ball (brain region) represents the degree of difference in active index between MDD group and HC group. Results show that SMN and VAN not only contain a larger number of brain regions, but these brain regions display very significant reductions in active index. These results confirm previous findings that abnormal FCs exist within and between regions in SMN and VAN of MDD group~\cite{javaheripour2021altered, yang2021disrupted, yan2020quantitative}. A recent study adopted independent component analysis (ICA) and Granger causality analysis (GCA) also revealed that MDD group had significantly decreased intra-FCs within VAN~\cite{luo2021abnormal}, which can be reasonably explained by our active index.




\section{Discussion \& Conclusion}
\label{sec:discuss}

In this paper, we propose the active index as an effective graph theory metric for brain functional analyses of MDD patients. We demonstrate the active index is more sensitive than both PC of diverse club and strength of rich club, which actually makes full use of the complementarity of diverse club and rich club in community-scale distributions. The disrupted active indexes across all brain regions may be a key manifestation that promotes the understanding of why brain networks of MDD patients can not work as economically as HC subjects. The combination of active index, diverse club, and rich club may be a future research direction to facilitate the functional analysis of other brain disorders or diseases.

\section{Compliance with Ethical Standards}
\label{sec:ethical}

This research study was conducted retrospectively using human subject data made available in open access by~\cite{yan2019reduced}. Ethical approval was not required as confirmed by the license attached with the open-access data.

\section{Acknowledgements}
\label{sec:acknowledgments}
This work was supported in part by the Zhejiang Provincial Innovation Team Project (No. 2020R01001) and in part by the Zhejiang Lab’s International Talent Fund for Young Professionals (ZJ2020JS013). Image data used in this study were provided by the members of REST-meta-MDD Consortium.




\bibliographystyle{IEEEbib}
\fontsize{9pt}{10pt} 
\selectfont
\bibliography{strings,refs}

\begin{thebibliography}{10}

\bibitem{yao2020morphological}
Zhijun Yao, Yu~Fu, Jianfeng Wu, Wenwen Zhang, Yue Yu, Zicheng Zhang, Xia Wu,
  Yalin Wang, and Bin Hu,
\newblock ``Morphological changes in subregions of hippocampus and amygdala in
  major depressive disorder patients,''
\newblock {\em Brain imaging and behavior}, vol. 14, no. 3, pp. 653--667, 2020.

\bibitem{yao2019structural}
Zhijun Yao, Ying Zou, Weihao Zheng, Zhe Zhang, Yuan Li, Yue Yu, Zicheng Zhang,
  Yu~Fu, Jie Shi, Wenwen Zhang, et~al.,
\newblock ``Structural alterations of the brain preceded functional alterations
  in major depressive disorder patients: evidence from multimodal
  connectivity,''
\newblock {\em Journal of affective disorders}, vol. 253, pp. 107--117, 2019.

\bibitem{sagar2020burden}
Rajesh Sagar, Rakhi Dandona, Gopalkrishna Gururaj, RS~Dhaliwal, Aditya Singh,
  Alize Ferrari, Tarun Dua, Atreyi Ganguli, Mathew Varghese, Joy~K Chakma,
  et~al.,
\newblock ``The burden of mental disorders across the states of india: the
  global burden of disease study 1990--2017,''
\newblock {\em The Lancet Psychiatry}, vol. 7, no. 2, pp. 148--161, 2020.

\bibitem{zhong2020prevalence}
Bao-Liang Zhong, Yi-Fan Ruan, Yan-Min Xu, Wen-Cai Chen, and Ling-Fei Liu,
\newblock ``Prevalence and recognition of depressive disorders among chinese
  older adults receiving primary care: a multi-center cross-sectional study,''
\newblock {\em Journal of affective disorders}, vol. 260, pp. 26--31, 2020.

\bibitem{bassett2017network}
Danielle~S Bassett and Olaf Sporns,
\newblock ``Network neuroscience,''
\newblock {\em Nature neuroscience}, vol. 20, no. 3, pp. 353--364, 2017.

\bibitem{bassett2018nature}
Danielle~S Bassett, Perry Zurn, and Joshua~I Gold,
\newblock ``On the nature and use of models in network neuroscience,''
\newblock {\em Nature Reviews Neuroscience}, vol. 19, no. 9, pp. 566--578,
  2018.

\bibitem{yu2020abnormal}
Zhinan Yu, Jiaolong Qin, Xinyuan Xiong, Fengguo Xu, Jun Wang, Fengzhen Hou, and
  Albert Yang,
\newblock ``Abnormal topology of brain functional networks in unipolar
  depression and bipolar disorder using optimal graph thresholding,''
\newblock {\em Progress in Neuro-Psychopharmacology and Biological Psychiatry},
  vol. 96, pp. 109758, 2020.

\bibitem{yun2021graph}
Je-Yeon Yun and Yong-Ku Kim,
\newblock ``Graph theory approach for the structural-functional brain
  connectome of depression,''
\newblock {\em Progress in Neuro-Psychopharmacology and Biological Psychiatry},
  vol. 111, pp. 110401, 2021.

\bibitem{liu2021disrupted}
Xinyi Liu, Cancan He, Dandan Fan, Yao Zhu, Feifei Zang, Qing Wang, Haisan
  Zhang, Zhijun Zhang, Hongxing Zhang, and Chunming Xie,
\newblock ``Disrupted rich-club network organization and individualized
  identification of patients with major depressive disorder,''
\newblock {\em Progress in Neuro-Psychopharmacology and Biological Psychiatry},
  vol. 108, pp. 110074, 2021.

\bibitem{bertolero2017diverse}
Max~A Bertolero, BT~Thomas Yeo, and Mark D’Esposito,
\newblock ``The diverse club,''
\newblock {\em Nature communications}, vol. 8, no. 1, pp. 1--11, 2017.

\bibitem{daianu2016disrupted}
Madelaine Daianu, Adam Mezher, Mario~F Mendez, Neda Jahanshad, Elvira~E
  Jimenez, and Paul~M Thompson,
\newblock ``Disrupted rich club network in behavioral variant frontotemporal
  dementia and early-onset a lzheimer's disease,''
\newblock {\em Human brain mapping}, vol. 37, no. 3, pp. 868--883, 2016.

\bibitem{verhelst2018impaired}
Helena Verhelst, Catharine Vander~Linden, Toon De~Pauw, Guy Vingerhoets, and
  Karen Caeyenberghs,
\newblock ``Impaired rich club and increased local connectivity in children
  with traumatic brain injury: Local support for the rich?,''
\newblock {\em Human brain mapping}, vol. 39, no. 7, pp. 2800--2811, 2018.

\bibitem{dosenbach2010prediction}
Nico~UF Dosenbach, Binyam Nardos, Alexander~L Cohen, Damien~A Fair, Jonathan~D
  Power, Jessica~A Church, Steven~M Nelson, Gagan~S Wig, Alecia~C Vogel,
  Christina~N Lessov-Schlaggar, et~al.,
\newblock ``Prediction of individual brain maturity using fmri,''
\newblock {\em Science}, vol. 329, no. 5997, pp. 1358--1361, 2010.

\bibitem{rosvall2008maps}
Martin Rosvall and Carl~T Bergstrom,
\newblock ``Maps of random walks on complex networks reveal community
  structure,''
\newblock {\em Proceedings of the national academy of sciences}, vol. 105, no.
  4, pp. 1118--1123, 2008.

\bibitem{van2011rich}
Martijn~P Van Den~Heuvel and Olaf Sporns,
\newblock ``Rich-club organization of the human connectome,''
\newblock {\em Journal of Neuroscience}, vol. 31, no. 44, pp. 15775--15786,
  2011.

\bibitem{yan2019reduced}
Chao-Gan Yan, Xiao Chen, Le~Li, Francisco~Xavier Castellanos, Tong-Jian Bai,
  Qi-Jing Bo, Jun Cao, Guan-Mao Chen, Ning-Xuan Chen, Wei Chen, et~al.,
\newblock ``Reduced default mode network functional connectivity in patients
  with recurrent major depressive disorder,''
\newblock {\em Proceedings of the National Academy of Sciences}, vol. 116, no.
  18, pp. 9078--9083, 2019.

\bibitem{javaheripour2021altered}
Nooshin Javaheripour, Meng Li, Tara Chand, Axel Krug, Tilo Kircher, Udo
  Dannlowski, Igor Nenadi{\'c}, J~Paul Hamilton, Matthew~D Sacchet, Ian~H
  Gotlib, et~al.,
\newblock ``Altered resting-state functional connectome in major depressive
  disorder: a mega-analysis from the psymri consortium,''
\newblock {\em Translational Psychiatry}, vol. 11, no. 1, pp. 1--9, 2021.

\bibitem{yang2021disrupted}
Hong Yang, Xiao Chen, Zuo-Bing Chen, Le~Li, Xue-Ying Li, Francisco~Xavier
  Castellanos, Tong-Jian Bai, Qi-Jing Bo, Jun Cao, Zhi-Kai Chang, et~al.,
\newblock ``Disrupted intrinsic functional brain topology in patients with
  major depressive disorder,''
\newblock {\em Molecular psychiatry}, pp. 1--9, 2021.

\bibitem{yan2020quantitative}
Baoyu Yan, Xiaopan Xu, Mengwan Liu, Kaizhong Zheng, Jian Liu, Jianming Li, Lei
  Wei, Binjie Zhang, Hongbing Lu, and Baojuan Li,
\newblock ``Quantitative identification of major depression based on
  resting-state dynamic functional connectivity: A machine learning approach,''
\newblock {\em Frontiers in neuroscience}, vol. 14, pp. 191, 2020.

\bibitem{luo2021abnormal}
Liang Luo, Huawang Wu, Jinping Xu, Fangfang Chen, Fengchun Wu, Chao Wang, and
  Jiaojian Wang,
\newblock ``Abnormal large-scale resting-state functional networks in drug-free
  major depressive disorder,''
\newblock {\em Brain imaging and behavior}, vol. 15, no. 1, pp. 96--106, 2021.

\end{thebibliography}

\end{document}